\documentclass[conference]{IEEEtran}
\usepackage[dvips]{color}
\usepackage{epsf}
\usepackage{times}
\usepackage{epsfig}
\usepackage{graphicx}
\usepackage{epstopdf}
\usepackage{pstricks}
\usepackage{amsmath}
\usepackage{amssymb}
\usepackage{amsxtra}
\usepackage{here}
\usepackage{rawfonts}
\usepackage{times}
\usepackage{url}
\usepackage{cite}
\usepackage{caption}
\usepackage{subcaption}
\usepackage{algorithm}
\usepackage{algorithmic}
\usepackage{breakurl}

\IEEEoverridecommandlockouts
\begin{document}
\title{Transfer Learning for Device Fingerprinting with Application to Cognitive Radio Networks\vspace{-0.5cm}}
\author{\IEEEauthorblockN{Yaman Sharaf-Dabbagh and Walid Saad}
\IEEEauthorblockA{Wireless@VT, Bradley Department of Electrical and Computer Engineering, Virginia Tech, VA, USA\\
Email: \{yamans, walids\}@vt.edu} \thanks{This research was supported by Virginia Tech's Institute for Critical Technology and Applied Science (ICTAS) and the U.S. National Science Foundation under Grant CNS-1524634.\vspace{-0.1cm}} \vspace{-0.6cm}}
\maketitle

\begin{abstract}
Primary user emulation~(PUE) attacks are an emerging threat to cognitive radio~(CR) networks in which malicious users imitate the primary users~(PUs) signals to limit the access of secondary users~(SUs). Ascertaining the identity of the devices is a key technical challenge that must be overcome to thwart the threat of PUE attacks. Typically, detection of PUE attacks is done by inspecting the signals coming from all the devices in the system, and then using these signals to form unique fingerprints for each device. Current detection and fingerprinting approaches require certain conditions to hold in order to effectively detect attackers. Such conditions include the need for a sufficient amount of fingerprint data for users or the existence of both the attacker and the victim PU within the same time frame. These conditions are necessary because current methods lack the ability to learn the behavior of both SUs and PUs with time. In this paper, a novel transfer learning~(TL) approach is proposed, in which abstract knowledge about PUs and SUs is transferred from past time frames to improve the detection process at future time frames. The proposed approach extracts a high level representation for the environment at every time frame. This high level information is accumulated to form an abstract knowledge database. The CR system then utilizes this database to accurately detect PUE attacks even if an insufficient amount of fingerprint data is available at the current time frame. The dynamic structure of the proposed approach uses the final detection decisions to update the abstract knowledge database for future runs. Simulation results show that the proposed method can improve the performance with an average of $3.5\%$ for only $10\%$ relevant information between the past knowledge and the current environment signals.
\end{abstract}
\IEEEpeerreviewmaketitle

\vspace{-0.1cm}
\section{Introduction}
Cognitive radio~(CR) is an emerging wireless paradigm that enables wireless devices, called secondary users~(SUs), to share the available spectrum whenever the licensed primary users~(PUs) are idle~\cite{akyildiz2006next}. However, this dynamic spectrum sharing can introduce many security threats, one of which is the primary user emulation~(PUE) attack~\cite{chen2008defense}. In PUE attacks, a malicious attacker emulates the PU signals to force legitimate SUs to vacate the spectrum even though no real PUs are currently present. This type of attacks significantly decreases the total performance of legitimate SUs and is detrimental to the overall CR system.

Many defense and detection strategies have been proposed to prevent PUE attacks in CR systems. These strategies fall into the following main categories: authentication and encryption based such as in~\cite{chandrashekar2010primary}, localization-based~\cite{chen2008defense}, transmit power estimation techniques~\cite{anand2008analytical}
, and smart strategies based on supervised machine learning techniques such as in~\cite{pu2011primary} and~\cite{yuan2012defeating} or game theory~\cite{li2011dogfight}. For PU authentication and encryption. the authors in~\cite{chandrashekar2010primary} proposed to add helper nodes to the system that authenticate PUs through link signatures, and hence inform SUs with available spectrum. Within the scope of localization-based techniques, the authors in~\cite{chen2008defense} exploited the received signal strength (RSS) distribution at different locations in the system for device fingerprinting. However, both the authentication and the localization-based techniques will require additional nodes for authentication or location reference, which can increase system complexity. Localization-based techniques also assume the PUs to be stationary with minimal channel variations, which is clearly not the case in practical CRs in which users constantly change their locations.
In~\cite{anand2008analytical}, the authors estimated the mean and the variance of the received power at both a licensed PU and a malicious user using Fenton approximation. These estimations are compared with the probability ratio test on the received signals to detect PUE attacks. However, these techniques assume that the transmission power levels of attackers are significantly different than the fixed known transmission power levels of the PUs. 
Using neural networks, the authors in~\cite{pu2011primary} developed an approach to distinguish between PU and PUE attacker. The use of belief propagation was proposed in~\cite{yuan2012defeating}, in which each SU calculates the belief based on the local and compatibility functions and the beliefs are exchanged between neighboring users and then used to detect PUE attacks. The work in~\cite{li2011dogfight} adopts a game-theoretic approach in which the PUE defense problem is a dog fight game between defending SUs and the attacker. 

Previous supervised machine learning based techniques such as in \cite{pu2011primary ,yuan2012defeating,li2011dogfight} require knowledge about users and authorized signals. However, this kind of information is not always available in real CR implementations.
Device fingerprinting~\cite{desmond2008identifying} is a process to extract device-specific information from the device transmitted signals. These fingerprints form signatures that uniquely identify each individual PU or SU in the system. Some of the features that can be used as fingerprints~\cite{ureten2007wireless} include the carrier frequency difference~(CFD), phase shift difference~(PSD), and second-order cyclostationary feature~(SOCF).
Since device fingerprinting requires no prior knowledge about PUs and SUs, unsupervised machine learning techniques can be used to detect PUE attacks. The authors in~\cite{nguyen2012identifying} used an unsupervised Bayesian mixture model to classify fingerprints into different groups. Each group represents a unique device in the system. A PUE attack is detected if two groups, physical devices, share the same ID, e.g. MAC address. Another benefit of the unsupervised techniques is that they are passive and require no additional hardware to be added to the system unlike localization based techniques. However, unsupervised fingerprinting techniques are applied at each time frame independently. Therefore, sufficient amount of fingerprints is required at each time frame, and it is required to have fingerprints for both the malicious attacker and the victim PU at the same time frame.

The main contribution of this paper is to propose a novel transfer learning based framework in which knowledge from past time frames is transferred to current time frame to improve the final PUE attacks detection decisions. To our knowledge, this paper is the \emph{first to develop a transfer learning approach for device fingerprinting}. Transferring knowledge is significant in fingerprinting scenarios in which an insufficient amount of fingerprints is available at the current time frame. Our proposed framework extracts abstract information about all SU and PU devices in the CR system. This abstract knowledge is used to boost the output of an unsupervised PUE attacks detection algorithm. The transfer learning part in the proposed framework updates the abstract knowledge after each time frame.
Extensive simulations demonstrates that the transfer approach improves the efficiency of PUE attacks detection especially in the case of insufficient availability of fingerprints.
Results show that using the proposed framework the performance is enhanced with an average of $3.5\%$ for only $10\%$ of relevant information between the past knowledge and the current environment signals.

The rest of this paper is organized as follows. Section~\ref{sec:model} describes our system model. In Section~\ref{sec:method} we explain in detail the proposed transfer learning framework. Simulation results and discussions are presented in Section~\ref{sec:results}. Finally, conclusions are drawn in Section~\ref{sec:end}.

\vspace{-0.1cm}
\section{System Model}\label{sec:model}
Consider a CR system composed of PUs and SUs, with the total number of PUs and SUs being $N$. Each user has a unique identifier~(ID), such as a MAC address. Secondary users~(SUs) can benefit from the spectrum only when all PUs are idle. Hence, SUs observe the signals of active PUs to avoid using the spectrum when it is used by a licensed PU. In this model, we consider a PUE attack scenario in which a malicious attacker copies the ID of one of the licensed PUs, and, in turn, forces the SUs to vacate the spectrum due to the fake PU-like signals sent from the attacker. The PUE attack causes a serious security threat and significantly degrades the total performance of the CR system.

Nevertheless, each device in the CR system, whether it is a PU or an SU, transmits signals that include device-specific information, such information is the carrier frequency difference~(CFD), phase shift difference~(PSD), and second-order cyclostationary feature~(SOCF). Such device-specific information is known to be unique to each transmitter~\cite{nguyen2012identifying}, and hence cannot be replicated by attackers.
The process to extract this device-specific information from the transmitted signals is called \emph{device fingerprinting}~\cite{desmond2008identifying}. In device fingerprinting, the extracted device-specific information, or fingerprints, forms signatures that uniquely identify each device in the system. Therefore, even though the PUE attacker shares the same ID with a real PU, the fingerprint data for the attacker is different from the fingerprint data for the attacked PU. Although our model is applicable to any device fingerprint features, we restrict our attention to the CFD, PSD, and SOCF.

To detect PUE attacks, a nonparametric Bayesian model can be used, such as the one in~\cite{nguyen2012identifying}. 
In this model, fingerprinting data is gathered from all the devices in the system and grouped into different clusters. 
Since fingerprints are unique for each device, each cluster of similar fingerprints belongs to one device only. Therefore, the number of clusters represents the real number of devices $N$ in the system.
The number of clusters is a key factor to detect the existence of a PUE attack. For example, if multiple clusters have the same ID, then a PUE attack has occurred.

To form the clusters, one can adopt the nonparametric approach proposed by Wood and Black~\cite{wood2008nonparametric}, that uses the infinite Gaussian mixture model (IGMM). In this model, the number of clusters, or mixtures, could be infinite. Using an IGMM, one can cluster a given set of data points, such as the PUs' and SUs' fingerprints, by estimating a number of Gaussian distributions, where each Gaussian distribution represent a cluster. The IGMM therefore requires a mixture input and a set of hyper-parameters $\beta$. These hyper-parameters include priors on the mean and variance of the multivariate Gaussian distributions. To integrate the hyper-parameters out of the IGMM, the mixture input needs to be generated from a distribution that is conjugate to the probability of the priors, $p(\mu_k,\sigma^2_k|\beta)$ with $\mu_k$ being the mean and $\sigma^2_k$ being the variance. Hence, the Dirichlet distribution is used for the mixing probability $p(\pi|\alpha)$ with $\pi$ being the mixing weights and $\alpha$ being the mixing parameter. The Dirichlet distribution is given by
\vspace{-0.1cm}\begin{align}
\text{Dir}(\alpha_1,\dots,\alpha_K) = \frac{1}{B(\alpha)}\prod\limits_{i=1}^K w_i^{\alpha_1-1},
\end{align}
for all $w_1,\dots,w_K>0$ where $w_1+\dots +w_K=1$. $B(\alpha)$ is the beta function, which can be expressed in terms of the gamma function as follows:
\vspace{-0.2cm}\begin{align}
B(\alpha) = \frac{\prod\nolimits_{i=1}^K \Gamma(\alpha_i)}{\Gamma(\sum\nolimits_{i=1}^K \alpha_i)}.
\end{align}
The Gibbs sampling process is then used to iteratively assign cluster labels to input data vectors, which in our case represent assigning fingerprints $f_i$ to devices. Gibbs sampler starts by initializing clusters for all the fingerprints $f_i$. Subsequently, the sampler removes the cluster label for one of the fingerprints and calculates its probability for new cluster assignment. The sampler iterates until convergence of all device assignments. Convergence is guaranteed   only for the case of infinite number of fingerprints. In the real case in which a finite number of fingerprints is available, a maximum number of iterations for the Gibbs sampler is chosen as a stopping criterion.

Such a nonparametric Bayesian method can be applied to a given set of fingerprints to detect PUE attacks. However, most such existing models, such as in~\cite{nguyen2012identifying}, are static and they apply this method independently at every time period. However, such static approaches require the CR system to have a sufficient amount of unique fingerprints for each device in the CR system to construct correct clusters for both the PUs and the SUs. 
In such a nonparametric system, there might not be a sufficient amount of fingerprint data for each device in the system. Therefore, there is a need to handle the case of having an insufficient amount of fingerprints to efficiently detect PUE attackers. To address this problem, next, we propose a transfer based framework in which fingerprinting results from past time frames are used to enhance the current PUE attacks detection.

\section{Proposed Transfer Learning Framework}\label{sec:method}
To enhance the fingerprinting process in a CR network, we use the powerful mathematical tool of transfer learning~\cite{pan2010survey}. In particular, we propose a new transfer learning based framework for device fingerprinting, as illustrated in Fig.~\ref{fig:method}.

The proposed framework consists of two main components. The first is the abstract knowledge database~(AKD), which is able to gather general information about devices in the CR system from all time frames. The second component is the transfer tool that updates the stored abstract knowledge database with the current time frame information.
The proposed framework receives the current fingerprint data from the environment, and suggests a clustering for the received fingerprint data based on the knowledge in the AKD.
The clustering decisions from the proposed framework are merged with the clustering decisions of the static nonparametric Bayesian model to form a final clustering decisions. These final clustering decisions represent the assignment of fingerprint data to devices in the CR system. The transfer tool updates the AKD with the final clustering decisions for future  PUE attacks detection.

Next, both the abstract knowledge database and the transfer learning algorithm will be described in detail.

\vspace{-0.1cm}
\subsection{Abstract Knowledge Database}
The abstract knowledge database~(AKD) includes general information about the devices in the CR system starting from time $t_0$, when the system was first deployed, and until the current time instant $t$. 
\begin{figure}[!t]
  \begin{center}
	\centering
 	   \includegraphics[width=6cm]{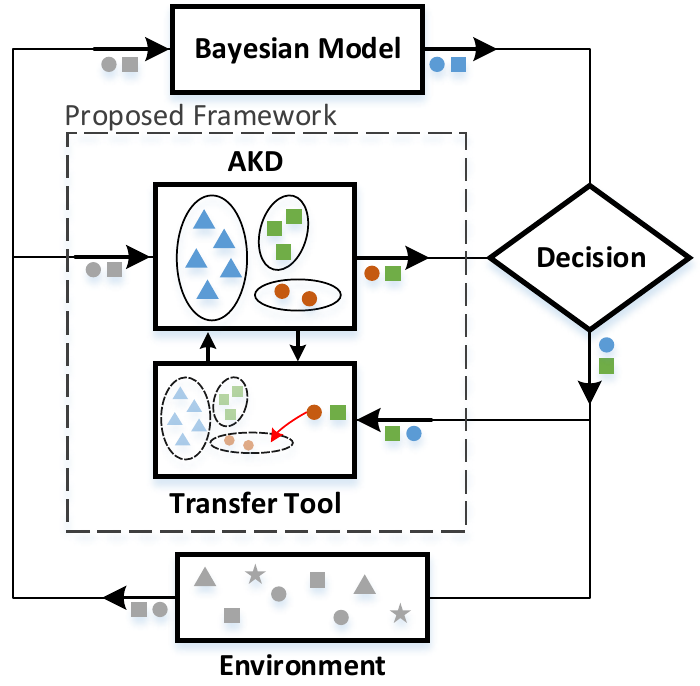}
  	  \caption{\small{\label{fig:method} Framework of the detection system with the proposed TL approach}}
  \end{center}
\end{figure}
The information stored in the AKD consists of a set of fingerprints $\mathcal{F}^{\textit{DB}}$, with \textit{DB} referring to the abstract knowledge database, grouped into multiple clusters $\{C^{\textit{DB}}_1,C^{\textit{DB}}_2,\dots,C^{\textit{DB}}_m\}$, where each cluster represents a device in the CR system. The AKD block in Fig.~\ref{fig:method} shows an example of 3 clusters of fingerprints.
Hence, each fingerprint $f_{i}^{\textit{DB}}\in \mathcal{F}^{\textit{DB}}$ has a corresponding cluster label $C^{\textit{DB}}_i\in \{C^{\textit{DB}}_1,C^{\textit{DB}}_2,\dots,C^{\textit{DB}}_m\}$ and a weight value $w_i$. 
Weights represent the amount of confidence in assigning fingerprint $f_{i}^{\textit{DB}}$ to cluster $C^{\textit{DB}}_i$. To reduce the number of fingerprints in the AKD, a similarity metric is introduced, which replaces two or more fingerprints $\{f_{i}^{\textit{DB}}$,$f_{\ell}^{\textit{DB}},\dots\}\in \mathcal{F}^{\textit{DB}}$ that are nearly identical with new fingerprint $f_{y}^{\textit{DB}}$, where $f_{y}^{\textit{DB}}$ minimizes the distances between $\{f_{i}^{\textit{DB}}$,$f_{\ell}^{\textit{DB}},\dots\}$:
\vspace{-0.1cm}\begin{align}
f_{y}^{\textit{DB}}\text{:} \textit{ }  \min_{f_{y}^{\textit{DB}}} \sum_{f_{x}^{\textit{DB}}\in \mathcal{X}} d\left( f_{y}^{\textit{DB}}, f_{x}^{\textit{DB}}\right),\label{eq:simi}
\end{align}
where $d$ is a distance measure and $\mathcal{X}$ is the set of similar fingerprints: $\mathcal{X}=\{f_{1}^{\textit{DB}},f_{2}^{\textit{DB}},\dots\}$ such that $\forall(f_{i}^{\textit{DB}},f_{\ell}^{\textit{DB}})\in~\mathcal{X}$, $d( f_{i}^{\textit{DB}},f_{\ell}^{\textit{DB}})\leq\epsilon$. Hence, the new set of fingerprints $\mathcal{F}^{\textit{DB}'}$ in the AKD equals to: $\mathcal{F}^{\textit{DB}'} = \left\{\mathcal{F}^{\textit{DB}}\cup\{f_{y}^{\textit{DB}} \}\right\}\setminus \{\mathcal{X}\}$.
For instance, if the distance measure is the l2-norm and we have two fingerprints $\{f_1^{\textit{DB}}, f_2^{\textit{DB}}\}$ with high similarity, $d( f_1^{\textit{DB}},f_2^{\textit{DB}})\leq\epsilon$ then the new fingerprint  $f_3^{\textit{DB}}$ using~(\ref{eq:simi}) will simply be the mean of both $\{f_1^{\textit{DB}}, f_2^{\textit{DB}}\}$. The distance threshold $\epsilon$ defines how close fingerprints need to be in order to be considered similar. Large values of $\epsilon$ allow more fingerprints to be merged and hence will result in a very generic abstract knowledge in the AKD. On the other hand, small values of $\epsilon$ tighten the similarity condition and result in a highly specific information in the AKD. 

When a new set of input fingerprint data $\mathcal{F}$ arrives from the environment, the goal is to group these fingerprints into clusters. 
In order to group the input fingerprints $\mathcal{F}$, 
the model compares these newly arriving fingerprints $\mathcal{F}$ with the fingerprints $\mathcal{F}^{\textit{DB}}$ stored in the AKD to find the set of 
fingerprints $f_{i}^{\textit{DB}}\subset~\mathcal{F}^{\textit{DB}}$ which is 
as close as possible to the input fingerprints $f_{\ell}\subset~\mathcal{F}$. 
Next, we use the clustering labels of $f_{i}^{\textit{DB}}$ as a suggested clustering labels for $f_{\ell}$. 
The model also uses the weights $w_i$ associated with each fingerprint $f_{i}^{\textit{DB}}$ in the AKD to provide a confidence level along with each suggested clustering label. A sigmoid logistic growth function is chosen to calculate the amount of confidence for each fingerprint in the database:
\vspace{-0.2cm}\begin{align}
\textit{CF$_i$} = \frac{1}{1+b\times e^{-c w_i}}\label{eq:conf},
\end{align}
where \textit{CF$_i$} is the confidence weight for the fingerprint $f_{i}^{\textit{DB}}$, $w_i$ is the weight associated with the fingerprint $f_{i}^{\textit{DB}}$ in the AKD, $b$ and $c$ are the parameters of the logistic function that control how steep the logistic function is which in turn affects the level of confidence assigned to each weight. The initial weights are chosen near the middle, based on the amount of confidence that we want to assign to the newly added fingerprints in the database.

\begin{algorithm}[!t]
\caption{Boosting labels of the Bayesian model}
\label{alg:alg1}
\begin{algorithmic}[1]
\REQUIRE Set of fingerprint data $\mathcal{F}$ from the environment
\ENSURE Set of labels $C^{\textit{DB}}_i$ and confidence values $\textit{CF}_{i}$ for all fingerprints $f_{i}\in \mathcal{F}$ 
\FOR{\textbf{each} fingerprint $f_{i}$ in the input}
	\FOR{\textbf{each} fingerprint $f_{j}^{\textit{DB}}$ in the AKD}
		\STATE Calculate similarity between $f_{i}$ and $f_{j}^{\textit{DB}}$
	\ENDFOR
	\STATE $\textit{idx} \leftarrow j$ (of the $f_{j}^{\textit{DB}}$  similar to $f_{i}$)
	\STATE $C^{\textit{DB}}_i \leftarrow$ labels of $f_{\textit{idx}}^{\textit{DB}}$
	\STATE $\textit{CF}_{i} = {1}/{[1+b\times \exp{(-c \times w_\textit{idx})}]}$
\ENDFOR
\RETURN clustering labels $C^{\textit{DB}}$, confidence \textit{CF}
\end{algorithmic}
\end{algorithm}
At specific time intervals, all the weights in the database are decreased by a fixed amount to reduce the confidence in old fingerprints which have not been used in the past several time frames. The parameters $b$ and $c$ in~(\ref{eq:conf}) determine how long it takes for a fingerprint to be considered old and be removed from the AKD.

Therefore, the output of our proposed framework is a set of suggested clustering labels for the input fingerprints $f_{i}$. A confidence weight is also provided along with each suggested clustering label.
The suggested clustering labels from our proposed framework boost the clustering decisions of the Bayesian model to construct final clustering labels. The boosting process pseudocode is shown in Algorithm~\ref{alg:alg1}.
\vspace{-0.1cm}
\subsection{Transfer Tool}
The purpose of the transfer tool is to use the final output decisions to update the AKD. 
At each time frame, both the 
Bayesian model and our proposed framework receive the current fingerprints $f_i$ from the environment, and group these fingerprints into clusters $C^{\textit{BY}}_i$ and $C^{\textit{DB}}_i$ respectively. 
The \textit{decision} block, shown in Fig.~\ref{fig:method}, merges the output of the Bayesian algorithm $(f_i,C^{\textit{BY}}_i)$ and the output of the proposed framework $(f_i,C^{\textit{DB}}_i)$ into a final output decision $(f_i,C_i)$, where $f_i$ is the input fingerprints, $C_i$ is the final cluster labels assignment.
Since the final decisions have different labels than the labels in the AKD, a mapping of labels is needed.
The transfer tool finds the right mapping of final decision labels $C_i$ to the labels in the AKD $C^{\textit{DB}}_i$ by solving:
\vspace{-0.2cm}\begin{align}
\min_{\textbf{\textit{M}}} \sum_{i=1}^{S} I\left(\textbf{\textit{M}}\left(C_i\right) ,C^{\textit{DB}}_i \right),\label{eq:map}
\end{align}
where $S$ is the total number of fingerprints in $\mathcal{F}$, $\textbf{\textit{M}}=[m_1,m_2,\dots,m_v]$ is the mapping vector between final cluster labels and cluster labels in the AKD such that $m_j\in\{1,2,\dots,V\}$, where $V$ is the number of labels, for any two elements $m_j$ and $m_k$ of  $\textbf{\textit{M}}, m_j\neq m_k$, $I$ is an indicator function in which $I=0$ when $\textbf{\textit{M}}\left(C_i\right)=C^{\textit{DB}}_i$ and $I=1$ when $\textbf{\textit{M}}\left(C_i\right)\neq C^{\textit{DB}}_i$.
For example, in the case of 2 clusters only, $\textbf{\textit{M}}$ can take two possible values: $\textbf{\textit{M}}=[1,2]$ or $[2,1]$.

After finding the right mapping of the labels from the final clustering output $C_i$ to the clustering labels $C^{\textit{DB}}_i$ in the AKD, the transfer tool updates the existing fingerprints $f^{\textit{DB}}_i$ in the AKD with the final output labels of the input fingerprints $f_{i}$. In the update process, the input fingerprint data $f_{i}$ is divided into two categories: fingerprints with high similarity to the fingerprints $f_{i}^{\textit{DB}}$ in the AKD, and fingerprints with low similarity to the
 fingerprints $f_{i}^{\textit{DB}}$ in the AKD. In the high similarity case, we increase the weights $w_i$ for all the fingerprints $f_{i}^{\textit{DB}}$ in the AKD that are of high similarity to 
 the input fingerprints $f_i$ to give these fingerprints a higher confidence weight. As for the fingerprints with low similarity, the input fingerprints $f_{i}$ are considered new to the AKD and are added to the AKD with initial associated weights $w_{init}$. The algorithm used to update fingerprints in the AKD is shown in Algorithm~\ref{alg:alg2}.
After specific time intervals, the weights for all the fingerprints in the database are decreased by $1$ to penalize for old unused fingerprints. All the fingerprints $f_{i}^{\textit{DB}}$ with weights below a certain threshold are removed from the AKD database.

\begin{algorithm}[!t]
\caption{Updating the AKD}
\label{alg:alg2}
\begin{algorithmic}[1]
\REQUIRE  Fingerprint data $f_{i}$ from the environment, and similarity levels between fingerprints $f_{i}$ from the environment and fingerprints $f_{j}^{\textit{DB}}$ from the AKD 
\ENSURE Updated AKD
\STATE Find the mapping of labels \textbf{\textit{M}} (between framework output labels and final output labels).
\FOR{\textbf{each} fingerprint $f_{i}$ with \textbf{\textit{high}} similarity to fingerprint $f_{j}^{\textit{DB}}$ in the AKD}
	\IF {$f_{i}$ label match the $f_{j}^{\textit{DB}}$}
		\STATE increase the weight of $f_{j}^{\textit{DB}}$ by $w_{step\_inc}$
	\ELSE 
		\STATE decrease the weight of $f_{j}^{\textit{DB}}$ by $w_{step\_dec}$
		\STATE add the new label mapping of $f_{i}$ to the AKD with initial weight $w_{init}$
	\ENDIF
\ENDFOR
\FOR{\textbf{each} fingerprint $f_{i}$ with \textbf{\textit{low}} similarity to fingerprint $f_{j}^{\textit{DB}}$ in the AKD}
	\STATE add the new fingerprint $f_{i}$ with its label to the AKD with initial weight $w_{init}$
\ENDFOR
\STATE decrease all weights for all fingerprints in the AKD by $1$ 
\end{algorithmic}
\end{algorithm}

\section{Simulation Results and Analysis}\label{sec:results}
For our simulations, we consider the nonparametric Bayesian method as our static PUE attacks detection model, and we evaluate the benefits of using our proposed transfer based framework to enhance the detection of the PUE attacks in the CR system. 
 \vspace{-0.2cm}
\subsection{Parameter Selection and Data Generation}
Even though the hyper-parameters $\beta$ were integrated out of the iterative Gibbs process as shown in Section~\ref{sec:model}, an appropriate choice of  hyper-parameters $\beta$ affects the speed of convergence of the Gibbs sampler. The hyper-parameters $\beta$ consist of priors to the $\mu_k$ and $\sigma_k$ of the multivariate Gaussian distributions. For example, The mean prior $\mu_0$ is usually set to be close to the mean value of all the fingerprints. The hyper-parameter $\kappa_0$, which is proportional to prior mean for $\sigma_k$, determines how likely the fingerprints are to be variant from each others. Therefore, for features like signal amplitude, a large value of $\kappa_0$ is chosen since signal amplitude feature has large variance between fingerprints. On the other hand, frequency features tend to have small variance between fingerprints and hence a small value of $\kappa_0$ is chosen for these features.
The Dirichlet distribution is used in the generation of the fingerprints for the simulations. Random means and variances are used for each label. 
\begin{figure*}[!t]
  \begin{center}
  \begin{subfigure}{.33\textwidth}
	\centering
 	   \includegraphics[width=6.6cm]{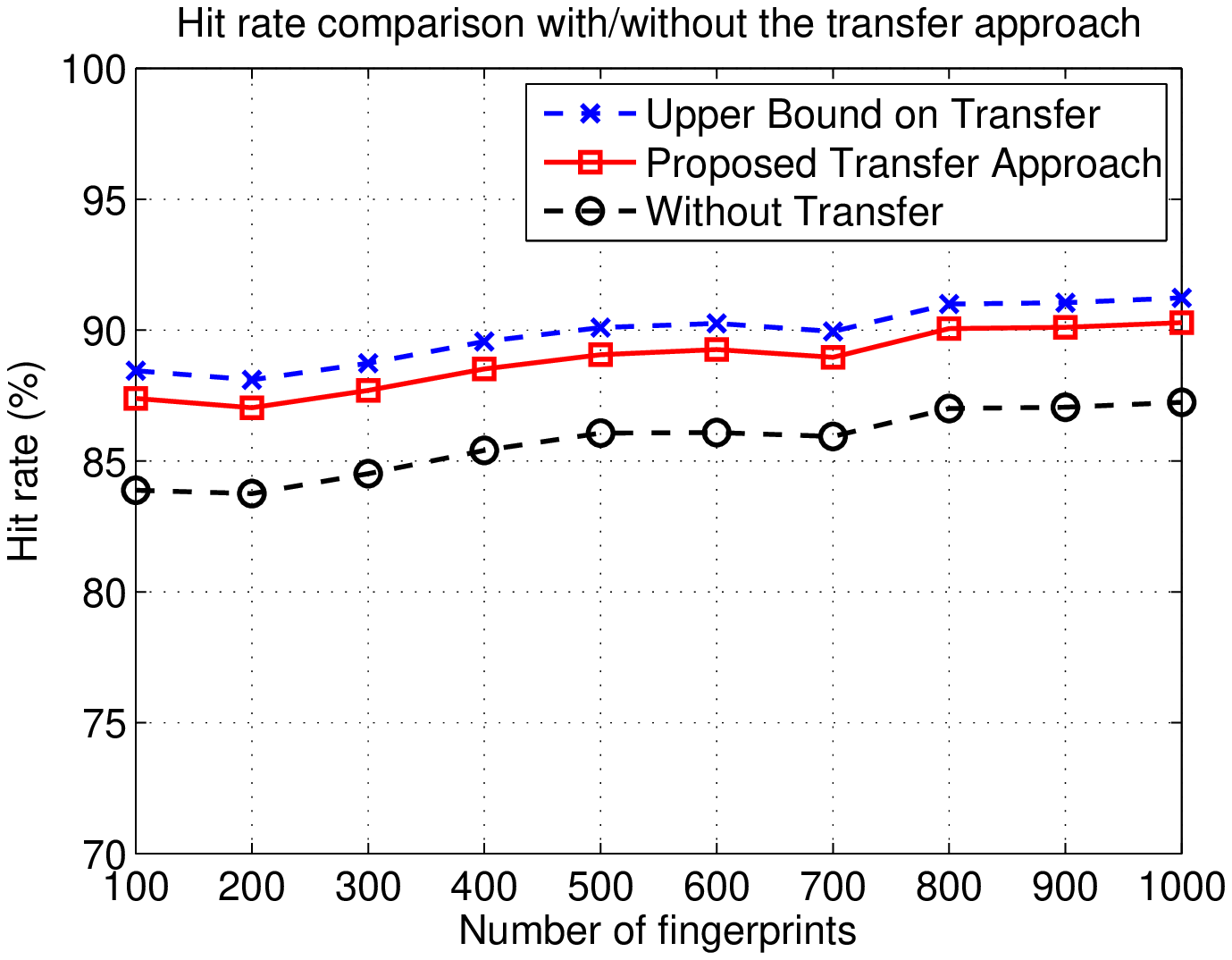}\vspace{-0.2cm}
  	  \caption{\label{fig:res1}}
  \end{subfigure}
  \begin{subfigure}{.33\textwidth}
	\centering
 	   \includegraphics[width=6.6cm]{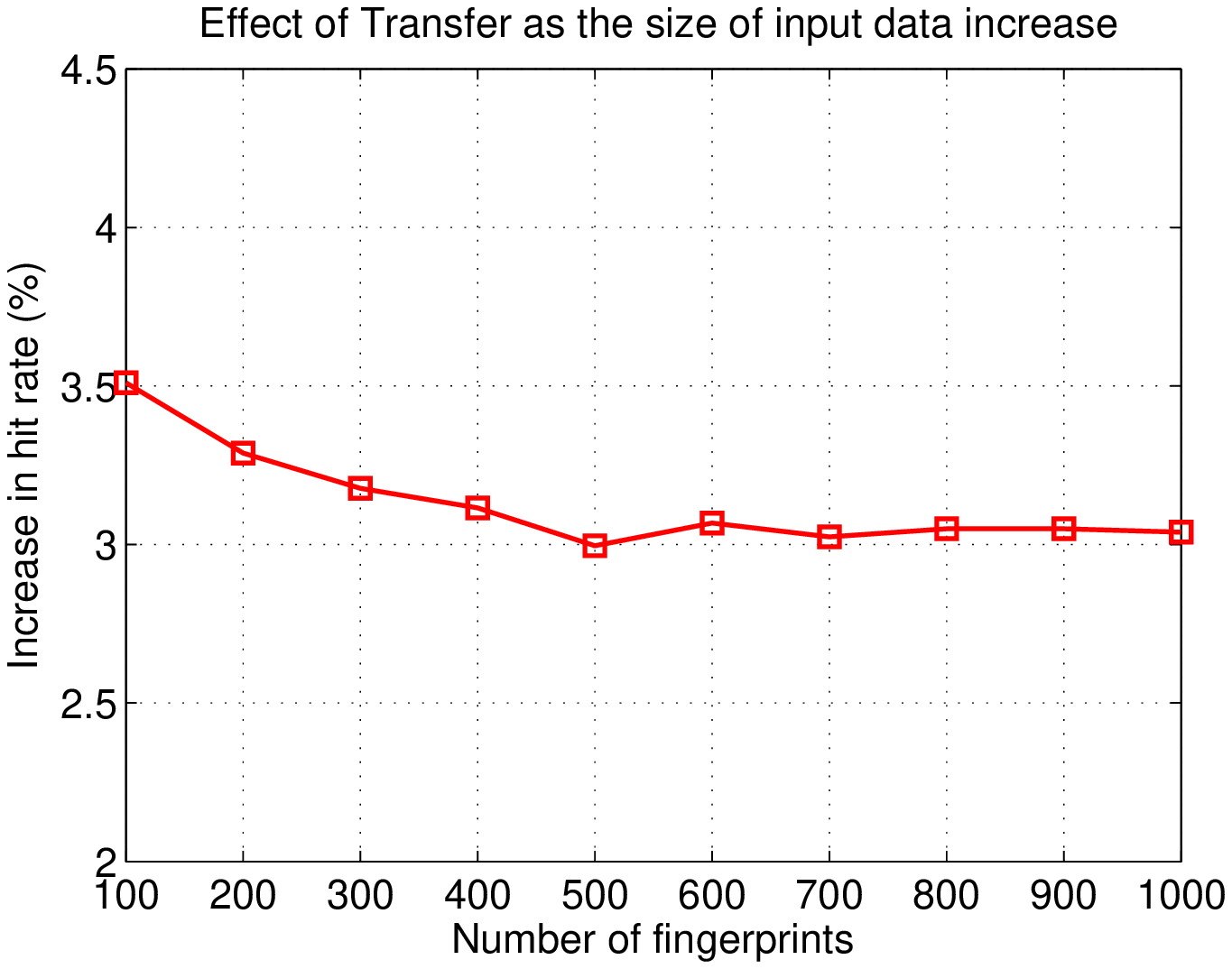}\vspace{-0.2cm}
  	  \caption{\label{fig:res2} }
	
  \end{subfigure}
  \begin{subfigure}{.32\textwidth}
	\centering
 	   \includegraphics[width=6.6cm]{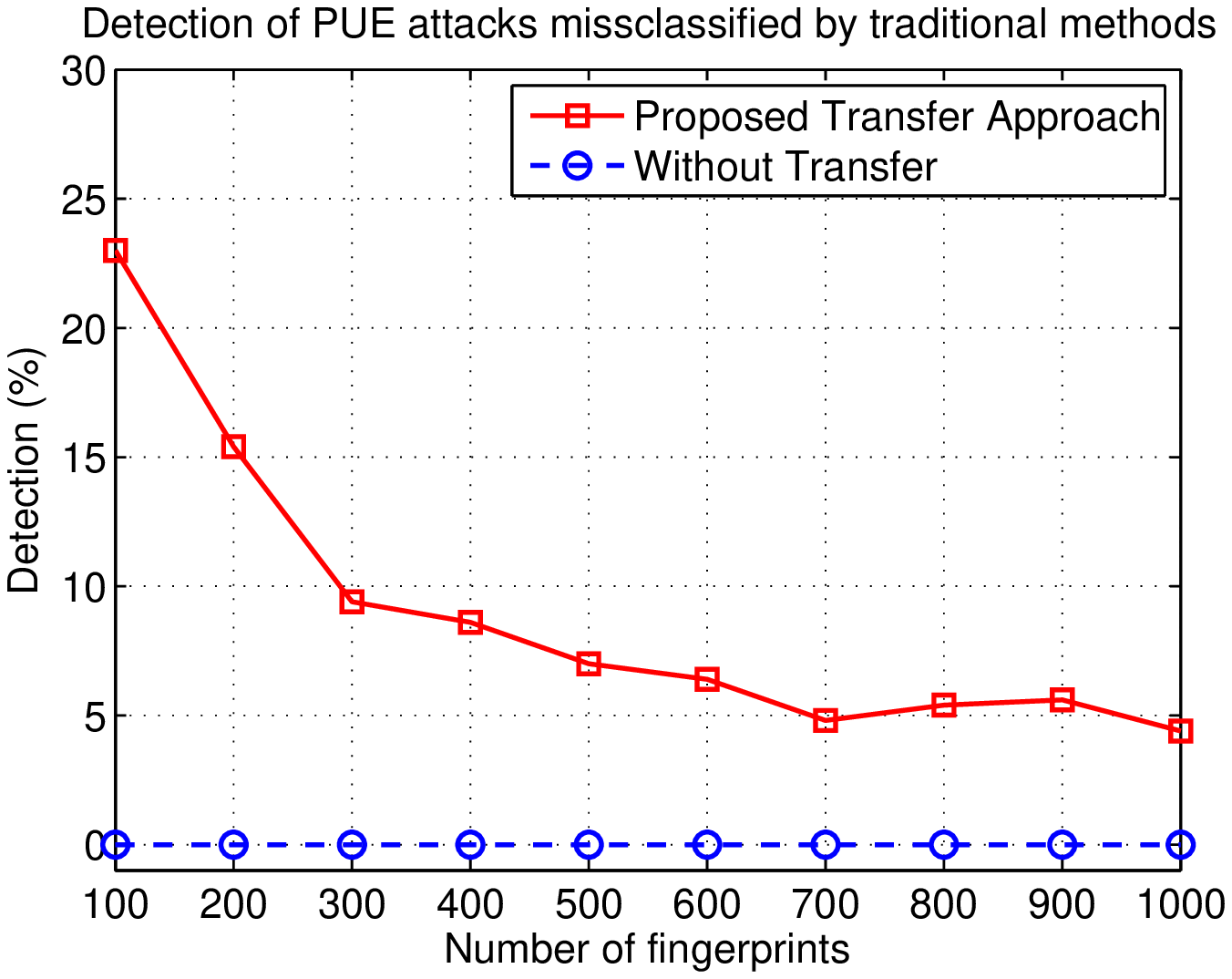}\vspace{-0.2cm}
  	  \caption{\label{fig:res3}}
  \end{subfigure}

  \caption{\small{\label{fig:both} Comparison between the proposed transfer learning approach and the traditional static approach (a) hit rate vs number of fingerprints (b) efficiency of using transfer learning vs number of fingerprints (c) percentage of PUE attacks detection vs number of fingerprints}}
  \end{center}

\vspace{-0.5cm}  
\end{figure*}
\vspace{-0.2cm}
\subsection{Experiments and Analysis}
In this section, we evaluate the performance of the nonparametric Bayesian model with and without our proposed transfer learning framework. We use two main metrics for evaluation: (i)~The hit rate of assigning the right label to each fingerprint, which is equals to the percentage of correctly assigned labels to fingerprints compared to the total number of fingerprints. (ii)~Missed attack rate, which is an indicator that missclassifying some fingerprints led into missclassifying a user in the system.


The second metric provides a better indicator on the performance enhancement due to using the transfer learning framework, this is due to the fact that even with high hit rate percentages, an attacker might not be detected if all the missclassified fingerprints belong to this attacker. On the other hand, even with low hit rate, accurate detection of attackers can be achieved if the missclassified fingerprints are distributed on various users.
The conducted experiments were repeated $500$ times and final results were averaged. The number of users chosen is $8$ with total number of fingerprints ranging from $100$ to $1000$ fingerprints. An upper bound on the transfer learning approach is shown in the plots, which presents the maximum hit rate possible with perfect transferred knowledge. This upper bound curve gives us an insight on the maximum performance that our proposed transfer approach can achieve.

In  Fig.~\ref{fig:res1}, we show the hit rate resulting from the different learning approaches as the number of fingerprints varies. From this figure, we can see that our proposed framework achieved an average enhancement of $3.5\%$ in total hit rate (for $100$ fingerprints) with only $10\%$ of transferable knowledge. Fig.~\ref{fig:res1} also shows that the transfer learning upper bound curve yields to highest output hit rate if all the $10\%$ transferable knowledge were completely used. The figure also shows that the maximum performance enhancement that can be reached through transfer learning is around $4.1\%$ making our model up to $75\%$ close to the optimal transfer case. As expected the more fingerprints we have that describes users the better the clusters are which yields to a higher total hit rate.

Fig.~\ref{fig:res2} shows the efficiency of using transfer learning as the number of fingerprints increases. As more fingerprints are available, there is less need to transfer knowledge and, thus, the amount of performance improvement resulting from the proposed approach will naturally decrease. This is due to the fact that having a large number of fingerprints will yield in a better clustering of labels by the detection algorithm which leaves small room of enhancement for the transfer approach.


In Fig.~\ref{fig:res3}, we show the number of times that the proposed transfer learning approach managed to correctly classify a PUE attacker that was missclassified in the traditional detection approach. As the number of fingerprints increases, missclassifying an attacker is less likely to happen. This is due to the fact that even if some fingerprints were missclassified, having other correctly classified fingerprints insures that the attacker is correctly detected. Fig.~\ref{fig:res3} shows that for $100$ fingerprints the proposed transfer approach detects $23\%$ of the times an attacker which was not detected in the traditional non-transfer approach.

\begin{figure}[!t]
  \begin{center}
	\centering
 	   \includegraphics[width=6.6cm]{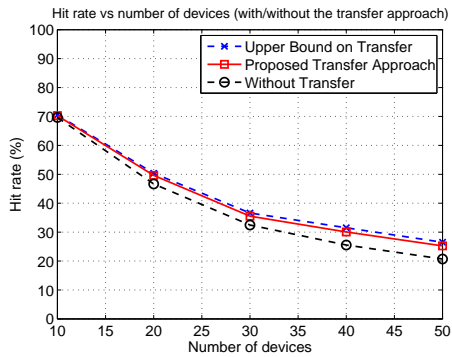}\vspace{-0.2cm}
  	  \caption{\small{\label{fig:res4}Hit rate comparison between the proposed and the traditional approaches with respect to the number of devices}}
  \end{center}

\end{figure}

Fig~\ref{fig:res4} shows the effect of increasing the number of devices on the total hit rate for a fixed number of fingerprints for each user. In this experiment, we obtain $25$ fingerprints for each user. As expected the hit rate decreases as the number of devices increases. For a fixed $10\%$ transferable information from past time frames, the figure shows that the effect of the proposed method increases as the number of devices increases, this is due to the fact that having 25 fingerprints for each user is  sufficient in the 10 users case, and as the number of users increases, more fingerprints are needed to correctly cluster users; and hence the more effective the transfer learning approach becomes. 

\vspace{-0.2cm}
\section{Conclusion}\label{sec:end}
In this paper, we have proposed a novel approach for performing device fingerprinting in wireless networks. In particular, we have introduced a novel framework, based on the machine learning tools of transfer learning, that enable static detection algorithms to benefit from past detection results. We have applied the framework for fingerprinting PUs and SUs in a cognitive radio network. Our results have shown that the proposed framework enhanced the performance with an average of $3.5\%$ for only $10\%$ relevant information between the past knowledge and the current environment fingerprints.

\vspace{0.2cm}
\def\baselinestretch{0.84}
{\bibliographystyle{IEEEtran}
\bibliography{paper}}

\end{document}